\documentclass[a4paper,12pt]{article}

\usepackage{ifpdf}

\newif\ifpdf
\ifx\pdfoutput\undefined
  \pdffalse
\else
  \pdfoutput=1
  \pdftrue
\fi

\RequirePackage{xspace} %
\RequirePackage{subfigure} %
\RequirePackage[centertags]{amsmath} %
\RequirePackage{amssymb}
\RequirePackage{wrapfig} %
\RequirePackage{calc} %
\RequirePackage{ifthen}
\RequirePackage{tabularx} %
\RequirePackage{flafter} %
\RequirePackage{fancyhdr} %

\ifpdf
  \RequirePackage[pdftex]{color}%
  \RequirePackage{colortbl}%
  \RequirePackage{array}%
  \RequirePackage[pdftex]{graphicx}

  \RequirePackage[ pdftex, plainpages = false, pdfpagelabels,
                 pdfpagelayout = useoutlines,
                 bookmarks,
                 breaklinks = true,
                 linktocpage,
                 pagebackref,                      % to include page numbers in bibliography
                 colorlinks = true,
                 linkcolor = blue,
                 urlcolor  = blue,
                 citecolor = blue,
                 anchorcolor = blue,
                 hyperindex = true,
                 hyperfigures
                 ]{hyperref}

\else
  \RequirePackage{color}
  \RequirePackage{colortbl}
   \RequirePackage{array}
  \RequirePackage[dvips]{graphicx}
  \RequirePackage{hyperref}
  \usepackage{rotating}
\fi

\usepackage{makeidx} % enable indexing
\usepackage{setspace} %enable setting spacing, %e.g. singlespace, onehalfspace, and doublespace
\usepackage{rotating} %enable sidewaysfigure
\usepackage{ecltree}
\usepackage{epic}
\usepackage{supertabular}  % this will allow the table not to break
\usepackage{color}
\usepackage{exscale}
\usepackage{fontenc}
\usepackage{ifthen}
\usepackage{latexsym}
\usepackage{makeidx}
\usepackage{syntonly}
\usepackage{inputenc}
\usepackage{graphicx}
\usepackage{setspace}
\usepackage{caption2}
\usepackage[english]{babel}
\usepackage[square, comma,numbers,sort&compress]{natbib}
\usepackage{hypernat}
\usepackage{boxedminipage}
\usepackage{framed}
\usepackage{longtable}
\usepackage[all]{hypcap}    %included to make the hyperlink go to the top of figure
\usepackage{algorithm2e}
\usepackage{algorithmic}
\usepackage{lscape}
\usepackage{pdflscape}

\newcommand{\note}[1]{\marginpar[left]{\singlespace \tiny #1}}

\newcommand{\ys}{yield-stress}

\newcommand{\Hs}      {\hspace{-0.5cm}} %
\newcommand{\sS}     {\tau}           % Shear stress
\newcommand{\wsS}    {\sS_{w}}        % Wall shear stress
\newcommand{\ysS}    {\sS_{o}}        % Yield stress
\newcommand{\CIF}     {\centering \includegraphics[width=2.7in]} %
\newcommand{\Vmin}    {\vspace{-0.2cm}} %

\renewcommand{\sectionmark}[1]%
      {\markright{\thesection\ #1}} %stops it capitalizing. #1 has value of section name

\renewcommand{\note}[1]{}

\doublespace %\onehalfspace

\title
{ %
\vspace*{3.0cm} \LARGE{\bf The Yield Condition in the Mobilization of Yield-Stress Materials in Distensible Tubes} \vspace*{4.0cm} \\
}

\author{Taha Sochi\footnote{University College London, Department of Physics \& Astronomy, Gower Street, London, WC1E 6BT.
Email: t.sochi@ucl.ac.uk.} \vspace*{5.0cm}}

\begin{document}

\maketitle %
\pagenumbering{arabic}

\newpage
\phantomsection \addcontentsline{toc}{section}{Contents} %
\tableofcontents

\newpage
\phantomsection \addcontentsline{toc}{section}{Abstract} \noindent
{\noindent \LARGE \bf Abstract} \vspace{0.5cm}\\
\noindent %

In this paper we investigate the yield condition in the mobilization of yield-stress materials in
distensible tubes. We discuss the two possibilities for modeling the yield-stress materials prior
to yield: solid-like materials and highly-viscous fluids and identify the logical consequences of
these two approaches on the yield condition. As part of this investigation we derive an analytical
expression for the pressure field inside a distensible tube with a Newtonian flow using a
one-dimensional Navier-Stokes flow model in conjunction with a pressure-area constitutive relation
based on elastic tube wall characteristics.


%%%%%%%%%%%%%%%%%%%%%%%%%%%%%%%%%%%  Head style  %%%%%%%%%%%%%%%%%%%%%%%%%%%%%%%%%%%
\pagestyle{headings} %
\addtolength{\headheight}{+1.6pt}
\lhead[{Chapter \thechapter \thepage}]%
      {{\bfseries\rightmark}}
\rhead[{\bfseries\leftmark}]%
     {{\bfseries\thepage}} %tell it to put page number at rhead
\headsep = 1.0cm               % Added 07 Sep 2006
%%%%%%%%%%%%%%%%%%%%%%%%%%%%%%%%%%%%%%%%%%%%%%%%%%%%%%%%%%%%%%%%%%%%%%%%%%%%%%%%%%%%

\newpage
%XXXXXXXXXXXXXXXXXXXXXXXXXXXXXXXXXXXXXXXXXXXXXXXXXXXXXXXXXXXXXXXXX
\section{Introduction}

Many naturally-occurring and synthetic materials are characterized by being \ys\ fluids, that is
they behave like solids under low shear stresses and as fluids on exceeding a minimum threshold
\ys\ \cite{BirdbookAH1987, CarreaubookKC1997, Barnes1999, SochiYield2010, SochiFeature2010}. Blood
and crude oils are some examples of naturally-occurring \ys\ fluids while many manufactured
products used in the food and petrochemical industries; like purees, yoghurts, slurries, drilling
muds, and polymeric solutions; are examples of synthetic \ys\ materials \cite{MerrillCP1969,
AlfarissP1984, WardhaughBT1988, MorrisRSGSB1989, AjienkaI1991, Barnes1999, HarteCC2007,
SochiArticle2010, LeeXNLS2011, DengThesis2012, PeinadoRHA2012, Livescu2012, CirielloF2012,
FarayolaOA2013, MahautGGLSM2013, SochiNonNewtBlood2013}. As these materials can reside and mobilize
in rigid conduits and structures, such as oil transportation pipelines and geological porous
formations, they can also reside and mobilize in compliant conduits and structures like blood
vessels and living porous tissues. Hence the investigation of the circumstances and conditions
under which these materials yield and start flowing in rigid and distensible containers is
important for modeling and analyzing many industrial and natural flow systems.

We are not aware of a previous attempt to find the condition for the yield of \ys\ materials in
distensible tubes. Past investigations are generally focused on the flow of \ys\ fluids in rigid
conduits and structures such as ensembles of conduits and porous media \cite{ChaplainMGC1992,
BalhoffT2004, ChenRY2005, SochiThesis2007, SochiB2008, SochiYield2010, BalanBNR2011, LiuYW2012,
ChevalierCCDCe2013, SochiYieldBal2013}. There are some other studies (e.g. \cite{VajraveluSDP2011})
that investigated the flow of \ys\ fluids in distensible conduits but the focus of these studies is
not on the yield condition but on the flow of these materials assuming they have already reached a
fluid phase state by satisfying the mobilization condition.

To clarify the purpose of the current investigation, we define the yield condition in general terms
that include rigid and distensible flow conduits as the minimum pressure drop across the conduit
that mobilizes the \ys\ material and forces it to flow following an initial solid state condition.
The yield point is therefore reached through increasing the pressure drop applied on the material
in its solid state gradually or suddenly. This definition is also valid in principle for the yield
condition in the bulk flow although this is not of interest to us in the current investigation.

In the present paper we investigate the yield condition in the mobilization of \ys\ materials
through distensible tubes using an elastic model for the tube wall that correlates the pressure at
a given axial coordinate to the corresponding cross sectional area. Although we assume a circular
cylindrical tube with a constant unstressed radius over its length, some of the presented arguments
at least can be generalized to include other cases although we are not intending to do so in the
current study.

In this investigation, we are only concerned with the yield condition as it is, without interest in
other issues related to the flow of \ys\ fluids in these conduits; hence any other dynamic issues
associated with the flow phase are out of scope of the present paper. Our plan for the paper is to
introduce the one-dimensional Navier-Stokes flow model which is widely used to describe the flow of
Newtonian fluids in distensible tubes. The reason for using this model, which is a Newtonian model
and not a \ys\ model, is that prior to yield the \ys\ materials according to one approach behave as
highly-viscous Newtonian fluids. We also discuss the pressure field inside the tube under such a
Navier-Stokes flow condition and how this field can be computed analytically and numerically
because it is needed for identifying the yield condition. This will be followed by a section in
which we investigate the yield condition in detail considering two approaches for modeling the \ys\
materials prior to yield. A general assessment of our proposed method for identifying the yield
condition with some synopsis conclusions will then follow.

%XXXXXXXXXXXXXXXXXXXXXXXXXXXXXXXXXXXXXXXXXXXXXXXXXXXXXXXXXXXXXXXXX
\section{One-Dimensional Navier-Stokes Model}\label{NavierStokes}

The flow of Newtonian fluids in a circular cylindrical tube with length $L$ and cross sectional
area $A$ assuming an incompressible laminar axi-symmetric flow with a negligible gravitational
force is described by the following one-dimensional Navier-Stokes system of mass and momentum
conservation principles
\begin{eqnarray}
\frac{\partial A}{\partial t}+\frac{\partial Q}{\partial x}&=&0\,\,\,\,\,\,\,\,\,\,\,\,\,
t\ge0,\,\,\, x\in[0,L]    \label{NSSystem1} \\
\frac{\partial Q}{\partial t}+\frac{\partial}{\partial x}\left(\frac{\alpha
Q^{2}}{A}\right)+\frac{A}{\rho}\frac{\partial p}{\partial
x}+\kappa\frac{Q}{A}&=&0\,\,\,\,\,\,\,\,\,\,\,\,\, t\ge0,\,\,\, x\in[0,L]     \label{NSSystem2}
\end{eqnarray}
where $Q$ is the volume flow rate, $x$ is the tube axial coordinate along its length, $t$ is the
elapsed time for the flow process, $\rho$ is the fluid mass density, $\alpha$ is the correction
factor for the momentum flux, $p$ is the axial pressure along the tube length which is a function
of the axial coordinate, and $\kappa$ is a viscosity friction coefficient which is normally defined
by $\kappa =\frac{2\pi\alpha\mu}{\rho(\alpha-1)}$ where $\mu$ is the fluid dynamic viscosity
\cite{BarnardHTV1966, SochiTechnical1D2013, SochiPois1DComp2013, SochiNavier2013}. In this context
we assume a no-slip at wall condition where the velocity of the fluid at the interface is identical
to the velocity of the solid \cite{SochiSlip2011, CostaWM2012, DamianouGM2013}. This Navier-Stokes
system is normally supported by a constitutive relation that links the cross sectional area at a
certain axial location to the corresponding axial pressure in a distensible tube, to close the
system in the three variables $Q$, $A$ and $p$ and hence provide a complete mathematical
description for such a flow in such a conduit.

The relation between the cross sectional area and the axial pressure in a distensible tube can be
described by many mathematical models depending on the characteristics of the tube wall and its
mechanical response to pressure such as being linear or non-linear, and elastic or viscoelastic.
The following is an example of a commonly used pressure-area constitutive elastic relation that
describes such a dependency

\begin{equation}\label{pAEq2}
p=\frac{\beta}{A_{o}}\left(\sqrt{A}-\sqrt{A_{o}}\right)
\end{equation}
where $A_{o}$ is the reference cross sectional area corresponding to the reference pressure which
in this equation is set to zero for convenience without affecting the generality of the results,
$A$ is the tube cross sectional area corresponding to the axial pressure $p$ as opposite to the
reference pressure, and $\beta$ is the tube wall stiffness coefficient which is usually defined by

\begin{equation}\label{beta}
    \beta = \frac{\sqrt{\pi}h_oE}{1-\varsigma^2}
\end{equation}
where $h_o$ is the tube wall thickness at the reference pressure, while $E$ and $\varsigma$ are
respectively the Young's elastic modulus and Poisson's ratio of the tube wall. In this context, we
assume a constant ambient pressure surrounding the tube that is set to zero and hence the reference
cross sectional area represents unstressed state with $A_{o}$ being constant over the axial
coordinate.

Now, to find the axial pressure field inside an elastic tube characterized by the mechanical
response of Equation \ref{pAEq2} we combine the pressure-area constitutive relation with the mass
and momentum equations of Navier-Stokes system. In the following we assume a pressure drop applied
on the tube by imposing two boundary conditions at $x=0$ and $x=L$ and hence it is identified by
two pressure boundary conditions, $p_{in}$ and $p_{ou}$, where

\begin{equation}
p_{in}\equiv p\left(x=0\right)>p_{ou}\equiv p\left(x=L\right)
\end{equation}
Another clarifying remark is that for a sustainable flow in a distensible tube, the tube axial
pressure should be a monotonically decreasing function of its axial coordinate. As a result, the
radius and the cross sectional area of the tube are also monotonically decreasing functions of the
axial coordinate. This fact is assumed in most of the following arguments although it may not be
stated explicitly.

From the pressure-area constitutive relation we have

\begin{equation}
p=\frac{\beta}{A_{o}}\left(\sqrt{A}-\sqrt{A_{o}}\right)
\end{equation}
and hence

\begin{equation}\label{ApEq}
A=\left(\frac{A_{o}}{\beta}p+\sqrt{A_{o}}\right)^{2}
\end{equation}
and

\begin{equation}
\frac{\partial A}{\partial
p}=2\frac{A_{o}}{\beta}\left(\frac{A_{o}}{\beta}p+\sqrt{A_{o}}\right)=\frac{2A_{o}^{2}}{\beta^{2}}p+\frac{2A_{o}^{3/2}}{\beta}
\end{equation}

For a steady state flow the time terms in the Navier-Stokes system are zero and hence from the
continuity equation (Equation \ref{NSSystem1}) $Q$ as a function of $x$ is constant. Consequently,
from the momentum equation of the Navier-Stokes system (Equation \ref{NSSystem2}) we obtain

\begin{equation}
\frac{\partial}{\partial x}\left(\frac{\alpha Q^{2}}{A}\right)+\frac{A}{\rho}\frac{\partial
p}{\partial x}+\kappa\frac{Q}{A}=0
\end{equation}
which can be manipulated to obtain the axial pressure gradient as follow

\begin{equation}
\alpha Q^{2}\frac{\partial}{\partial A}\left(\frac{1}{A}\right)\frac{\partial A}{\partial
p}\frac{\partial p}{\partial x}+\frac{A}{\rho}\frac{\partial p}{\partial x}+\kappa\frac{Q}{A}=0
\end{equation}

\begin{equation}
-\frac{\alpha Q^{2}}{A^{2}}\frac{\partial A}{\partial p}\frac{\partial p}{\partial
x}+\frac{A}{\rho}\frac{\partial p}{\partial x}+\kappa\frac{Q}{A}=0
\end{equation}

\begin{equation}
-\frac{\alpha
Q^{2}}{A^{2}}\left(\frac{2A_{o}^{2}}{\beta^{2}}p+\frac{2A_{o}^{3/2}}{\beta}\right)\frac{\partial
p}{\partial x}+\frac{A}{\rho}\frac{\partial p}{\partial x}+\kappa\frac{Q}{A}=0
\end{equation}

\begin{equation}
\frac{\partial p}{\partial x}\left[-\frac{\alpha
Q^{2}}{A^{2}}\left(\frac{2A_{o}^{2}}{\beta^{2}}p+\frac{2A_{o}^{3/2}}{\beta}\right)+\frac{A}{\rho}\right]=-\kappa\frac{Q}{A}
\end{equation}
that is

\begin{equation}
\frac{\partial p}{\partial x}=\frac{\kappa\frac{Q}{A}}{\frac{\alpha
Q^{2}}{A^{2}}\left(\frac{2A_{o}^{2}}{\beta^{2}}p+\frac{2A_{o}^{3/2}}{\beta}\right)-\frac{A}{\rho}}
\end{equation}

The last expression can be simplified to

\begin{equation}\label{dpdx}
\frac{\partial p}{\partial x}=\frac{\kappa
Q\left(\frac{A_{o}}{\beta}p+\sqrt{A_{o}}\right)^{2}}{\alpha
Q^{2}\left(\frac{2A_{o}^{2}}{\beta^{2}}p+\frac{2A_{o}^{3/2}}{\beta}\right)-\frac{\left(\frac{A_{o}}{\beta}p+\sqrt{A_{o}}\right)^{6}}{\rho}}
\end{equation}
which expresses the axial pressure gradient, $\frac{\partial p}{\partial x}$, as a function of the
tube axial pressure, $p$, only. For a one-dimensional steady state flow, the pressure is dependent
on the tube axial coordinate only and hence the partial derivative of Equation \ref{dpdx} can be
replaced with a total derivative. On separating the variables and carrying out the integration,
where $0\le x \le L$ and $p_{ou}\le p \le p_{in}$, we obtain

\begin{equation}\label{pAnalEq}
x=\frac{2\alpha
Q\ln\left(\frac{A_{o}}{\beta}p+\sqrt{A_{o}}\right)}{\kappa}-\frac{\beta\left(\frac{A_{o}}{\beta}p+\sqrt{A_{o}}\right)^{5}}{5\rho\kappa
QA_{o}}+C
\end{equation}
where $C$ is the constant of integration which can be determined from the boundary conditions. At
$x=0$ $p=p_{in}$, and hence

\begin{equation}
C=-\frac{2\alpha
Q\ln\left(\frac{A_{o}}{\beta}p_{in}+\sqrt{A_{o}}\right)}{\kappa}+\frac{\beta\left(\frac{A_{o}}{\beta}p_{in}+\sqrt{A_{o}}\right)^{5}}{5\rho\kappa
QA_{o}}
\end{equation}

The analytical solution of the pressure field, as given by Equation \ref{pAnalEq}, defines the
axial pressure field $p(x)$ as an implicit function of the axial coordinate $x$. Obtaining the
pressure field then requires either the employment of a simple numerical solver, based for example
on a bisection method, or changing the role of the independent and dependent variables and hence
obtaining $x$ as a function of $p$ where the value of pressure is constrained by the two boundary
conditions. The latter scheme, which is more convenient to use, leads to the same result as the
first scheme although the defining $x$-$p$ points are determined by the pressure points and hence
obtaining a smoothly defined pressure field may require computing more points than is needed from
the first scheme. In addition to the convenience, the solution obtained from the second scheme is
generally more accurate because it produces exact solutions within the available computing
precision although this extra accuracy may be of little value in practical circumstances.

\begin{figure}
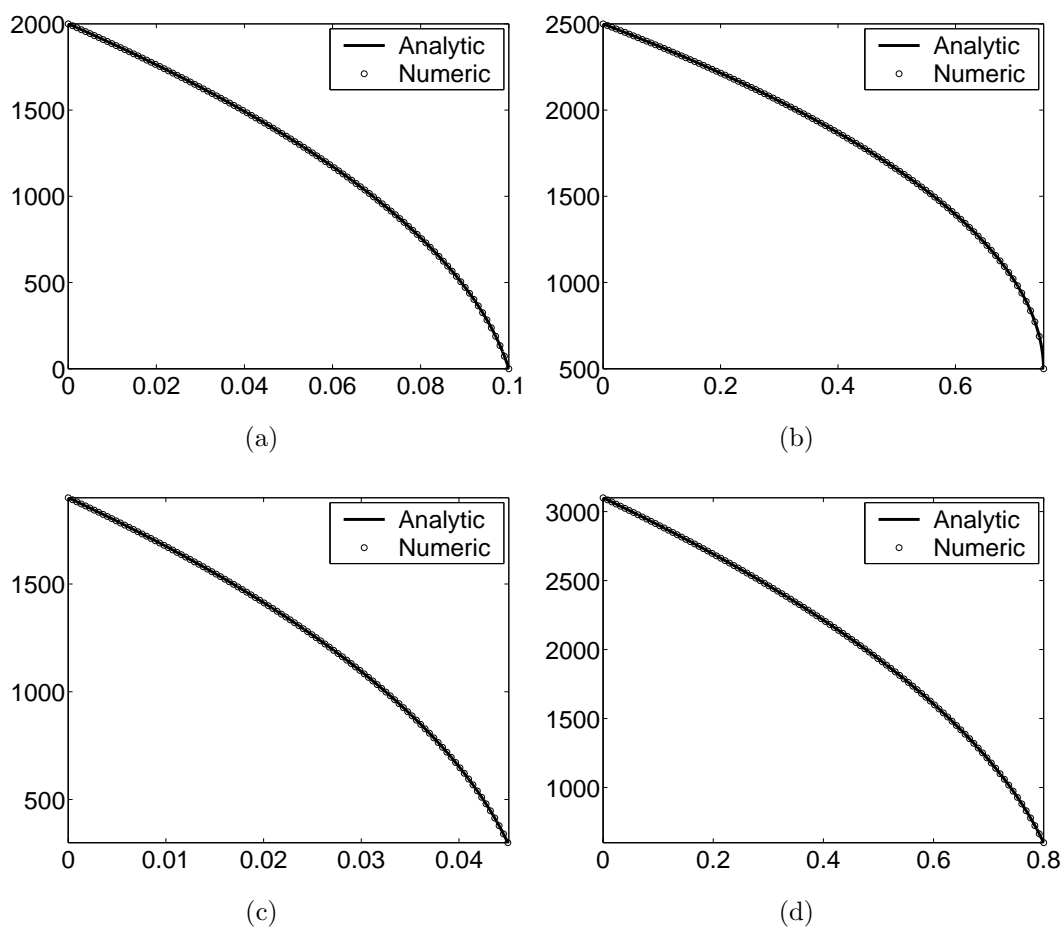

\centering %
\subfigure[]%
{\begin{minipage}[b]{0.5\textwidth} \CIF {g/pAnalFig1}
\end{minipage}}
\Hs
\subfigure[]%
{\begin{minipage}[b]{0.5\textwidth} \CIF {g/pAnalFig2}
\end{minipage}} \Vmin
%XXXXXXXXXXXXXXXXXXXXXXXXXXXXXXXXXXXXXXXXXXXXXXXXXXXXXXXXXXXXXXXXXXX
\centering %
\subfigure[]%
{\begin{minipage}[b]{0.5\textwidth} \CIF {g/pAnalFig3}
\end{minipage}}
\Hs
\subfigure[]%
{\begin{minipage}[b]{0.5\textwidth} \CIF {g/pAnalFig4}
\end{minipage}}
\caption{Comparing the analytical solution of the axial pressure field as given by Equation
\ref{pAnalEq} to the numerical solution from the residual-based lubrication approximation for the
Navier-Stokes flow in elastic tubes with (a) $R=0.01$, $L=0.1$, $\alpha=\frac{4}{3}$, $\beta=39.5$,
$\rho=900$, $\mu=0.01$, $p_{in}=2000$, $p_{ou}=0$, and $Q=2.36708\times 10^{-3}$, (b) $R=0.05$,
$L=0.75$, $\alpha=1.25$, $\beta=88.9$, $\rho=1300$, $\mu=0.025$, $p_{in}=2500$, $p_{ou}=500$, and
$Q=7.34585\times 10^{-2}$, (c) $R=0.003$, $L=0.045$, $\alpha=1.5$, $\beta=12.9$, $\rho=1100$,
$\mu=0.015$, $p_{in}=1900$, $p_{ou}=300$, and $Q=4.66119\times 10^{-5}$, and (d) $R=0.1$, $L=0.9$,
$\alpha=1.3$, $\beta=227.7$, $\rho=800$, $\mu=0.035$, $p_{in}=3100$, $p_{ou}=600$, and
$Q=2.49237\times 10^{-1}$. All the dimensional quantities are in standard SI units. In all four
sub-figures, the vertical axis represents the tube axial pressure in pascals while the horizontal
axis represents the tube axial coordinate in meters. \label{pAnalFig}}
\end{figure}

The analytical solution of Equation \ref{pAnalEq} was tested and validated by numerical solutions
from many typical flow examples using the lubrication approximation with a residual-based
Newton-Raphson solution scheme. This numerical method for obtaining the pressure field and flow
rate is based on discretizing the flow conduit into thin rings which are treated as an ensemble of
serially connected tubes over which a mass conserving characteristic flow (in this case the
one-dimensional Navier-Stokes flow in elastic cylindrical tubes) is sought by forming a system of
simultaneous equations derived from the boundary conditions for the boundary nodes and the
continuity of volumetric flow rate for the internal nodes. This system is then solved numerically
by an iterative non-linear solution scheme, such as Newton-Raphson method, to obtain the axial
pressure field inside the conduit, defined at the boundary and internal nodes, as well as the
mass-conserving flow rate within a given error tolerance. This numerical solving scheme for
obtaining the pressure field and flow rate is fully described in \cite{SochiPoreScaleElastic2013,
SochiConDivElastic2013, SochiCoDiNonNewt2013}. A sample of the results comparing the analytical to
the numerical solutions for some typical examples with a range of tube, flow and fluid parameters
are given in Figure \ref{pAnalFig}.

\clearpage

\clearpage
%XXXXXXXXXXXXXXXXXXXXXXXXXXXXXXXXXXXXXXXXXXXXXXXXXXXXXXXXXXXXXXXXX
\section{Yield Condition}

A general definition has been given earlier for the yield condition. However, since the pressure
field inside a distensible tube and the resulting flow rate are dependent on the actual value of
the two boundary conditions and not only on their difference (i.e. pressure drop)
\cite{SochiElastic2013} we define the yield condition for distensible tubes as the minimum inlet
pressure for a given outlet pressure that mobilizes the \ys\ material and initiates a measurable
flow. The generally accepted condition for the yield of a \ys\ material in a tube is given by
equating the magnitude of the wall shear stress $|\tau_{w}|$ to the \ys\ of the fluid $\tau_{o}$
where $\tau_{w}$ is defined as the ratio of the force normal to the tube axis $F_{\perp}$ to the
area of the luminal surface parallel to this force $A_{\parallel}$.

Our proposal for the condition that should be satisfied to reach the yield point is that yield
occurs {\em iff} all cross sectional points in the tube along its length reached their yield
condition simultaneously. This will only occur if the bottleneck in the tube reaches its yield
condition where the bottleneck is the location along the tube axis with the lowest magnitude of
wall shear stress. It is obvious that this is a necessary and sufficient condition for yield to
occur because if the bottleneck reached its yield condition with its wall shear stress just
exceeding the \ys\ then all other points along the tube axis should also satisfy the yield
condition.

For a rigid tube with a constant radius along its axial direction, $\tau_{w}$ is given by

\begin{equation}
\tau_{w}\equiv\frac{F_{\perp}}{A_{\parallel}}=\frac{\pi R^{2}\Delta P}{2\pi RL}=\frac{R\Delta
P}{2L}
\end{equation}
where $R$ and $L$ are the tube radius and length respectively while $\Delta P$ is the pressure drop
across the tube.

For a tube with a variable radius along its axial direction, including a distensible tube subjected
to a pressure gradient along its axis, the wall shear stress is a function of the tube axial
coordinate $x$ and hence the wall shear stress is defined locally as a function of $x$ by using
infinitesimal quantities of pressure drop, $\delta p$, and length, $\delta x$, belonging to a thin
cross sectional ring over which the change in radius is negligible, that is

\begin{equation}
\tau_{w}\left(x\right)\equiv\frac{F_{\perp}}{A_{\parallel}}\simeq\frac{\pi R^{2}\delta p}{2\pi
R\delta x}=\frac{R}{2}\frac{\delta p}{\delta x}
\end{equation}
which in the limit becomes

\begin{equation}
\tau_{w}\left(x\right)=\frac{R}{2}\frac{dp}{dx}
\end{equation}

Therefore the condition for the transition from the solid state to fluid phase, corresponding to
the yield and flow initiation, is defined by the following condition

\begin{equation}
|\tau_{w}|=\tau_{o}\,\,\,\,\,\,\,\,\,\Rightarrow\,\,\,\,\,\,\,\,\,\frac{R(x)}{2}\left|\frac{dp}{dx}\right|=\tau_{o}\label{YieldCondVar}
\end{equation}

There are two possibilities for modeling the \ys\ materials before reaching their yield point:
either they are solid-like materials or they are highly-viscous fluids. A detailed discussion about
this issue, among other relevant issues, is given in \cite{SochiYieldBal2013}. Our proposal for
identifying the yield point according to each one of these two possibilities is discussed in the
following two subsections.

%XXXXXXXXXXXXXXXXXXXXXXXXXXXX
\subsection{Solid-Like Materials}

According to this approach, the \ys\ materials before reaching their yield point behave like
solids. Hence, a logical assumption about the pressure field configuration inside the tube prior to
mobilization is to assume a linear pressure drop and hence a constant pressure gradient. In this
case the bottleneck occurs at the outlet boundary point because the radius of the tube is
monotonically decreasing and hence the outlet point is the one with the smallest radius which
implies that the wall shear stress as given by Equation \ref{YieldCondVar} is the lowest since
$\frac{dp}{dx}$ is assumed constant along the tube axis. Therefore the yield condition according to
the solid-like modeling approach is given by

\begin{equation}
\frac{R(L)}{2}\left|\frac{dp}{dx}\right|=\tau_{o}
\end{equation}
where $R(L)$ is the tube radius at the outlet boundary corresponding to $x=L$.

%XXXXXXXXXXXXXXXXXXXXXXXXXXXX
\subsection{Highly-Viscous Fluids}

According to this approach, the \ys\ materials before reaching their yield point are highly-viscous
fluids and hence they flow but with an infinitesimally small flow rate. Our proposal in this case
is that the material prior to yield should be modeled as a Newtonian fluid with a very high
viscosity. The assumption of a Newtonian fluid is justified by the fact that at such regimes of
very low rate of deformation the material is at its low Newtonian plateau since all non-Newtonian
rheological effects are induced only by a measurable deformation \cite{SochiYieldBal2013}. We
therefore use the one-dimensional Navier-Stokes flow model to identify the pressure field prior to
yield and hence the yield condition. This one-dimensional Navier-Stokes flow model is fully
described in \cite{SochiTechnical1D2013, SochiPois1DComp2013}. It is also outlined in section
\ref{NavierStokes} with a derivation of an analytical expression for the axial pressure as an
implicit function of the tube axial coordinate. The determination of the axial pressure field is
obviously needed for the identification of the yield condition because both $R$ and $\frac{dp}{dx}$
are dependent on the pressure field.

Our general approach for identifying the yield condition assuming a highly-viscous fluid prior to
yield is to test the axial pressure field as determined by the given inlet and outlet boundary
conditions. The purpose of this test is to identify the bottleneck by determining the lowest wall
shear stress in magnitude along the tube and hence verifying if the bottleneck has reached its
minimum threshold which equals the \ys\ value, $\tau_o$. As seen in Equation \ref{YieldCondVar},
$|\tau_w|$ is dependent on the pressure gradient and the tube radius as functions of the axial
coordinate. Hence a knowledge of these two quantities is required.

There are two ways for obtaining the pressure gradient and the tube radius. The first way is to
solve the pressure field numerically using the aforementioned residual-based lubrication approach.
The axial pressure gradient is then calculated from the numerical solution of the pressure field
{\it either} analytically from Equation \ref{dpdx} {\it or} numerically using for instance a finite
difference method. On knowing the axial pressure, the tube radius as a function of the axial
coordinate can also be computed from the pressure-area constitutive relation (Equation \ref{ApEq}).
The second way is to use the analytical solution of the pressure field, which is derived in section
\ref{NavierStokes}, and hence evaluate the axial pressure gradient and axial radius as in the case
of numerical solution of the pressure field, although the analytical evaluation of the pressure
gradient from Equation \ref{dpdx} may not be possible unless $Q$ is obtained from another method.
Both of these ways practically produce the same result if a sufficiently fine mesh is used for
discretizing the tube in the residual-based lubrication technique.

As indicated early, in the case of using the analytical expression of Equation \ref{pAnalEq} a
simple numeric solver, based for instance on a bisection method, can be used to evaluate the axial
pressure which is given as an implicit function of the tube axial coordinate. An alternative and
more convenient way that does not require the employment of a numerical solver is to use Equation
\ref{pAnalEq} to find the axial coordinate $x$ as a function of the axial pressure $p$, which is as
good for evaluating the axial pressure field as the other way round. The use of one of these
methods or the other (i.e. numerical or analytical with or without the employment of a numerical
solver) is a matter of choice and convenience. All our numerical experiments produced essentially
the same results on using these different methods and techniques.

However, it can be shown that the bottleneck, according to the highly-viscous fluid approach, is
always at the inlet boundary and hence the use of the numerical and analytical methods is
unnecessary apart from providing the one-sided pressure gradient at the inlet since the inlet
radius is already known from the inlet boundary condition. The justification of this is explained
in the following paragraphs.

As stated before, for a steady state flow the time terms in the Navier-Stokes system are zero and
hence from the continuity equation (Equation \ref{NSSystem1}) $Q$ as a function of $x$ is constant.
Hence, from the momentum equation of the Navier-Stokes system (Equation \ref{NSSystem2}) we obtain

\begin{equation}
\frac{\partial}{\partial x}\left(\frac{\alpha Q^{2}}{A}\right)+\frac{A}{\rho}\frac{\partial
p}{\partial x}+\kappa\frac{Q}{A}=0
\end{equation}

\begin{equation}
\frac{\partial p}{\partial x}=-\frac{\rho}{A}\left[\kappa\frac{Q}{A}+\frac{\partial}{\partial
x}\left(\frac{\alpha Q^{2}}{A}\right)\right]
\end{equation}

\begin{equation}
\frac{\partial p}{\partial x}=-\left(\frac{\kappa\rho Q}{A^{2}}+\frac{\rho\alpha
Q^{2}}{A}\frac{\partial A^{-1}}{\partial x}\right)
\end{equation}

\begin{equation}
\frac{\partial p}{\partial x}=-\left(\frac{\kappa\rho Q}{A^{2}}+\frac{\rho\alpha
Q^{2}}{A}\frac{\partial A^{-1}}{\partial A}\frac{\partial A}{\partial x}\right)
\end{equation}
that is

\begin{equation}
\frac{\partial p}{\partial x}=-\left(\frac{\kappa\rho Q}{A^{2}}-\frac{\rho\alpha
Q^{2}}{A^{3}}\frac{\partial A}{\partial x}\right)
\end{equation}

Now since $A$ is a monotonically decreasing function of $x$

\begin{equation}
-\frac{\partial A}{\partial x}=\left|\frac{\partial A}{\partial x}\right|
\end{equation}
and hence

\begin{equation}
\frac{\partial p}{\partial x}=-\left(\frac{\kappa\rho Q}{A^{2}}+\frac{\rho\alpha
Q^{2}}{A^{3}}\left|\frac{\partial A}{\partial x}\right|\right)
\end{equation}

Therefore

\begin{equation}
|\tau_{w}|=\frac{R}{2}\left|\frac{\partial p}{\partial x}\right|=\frac{R}{2}\left(\frac{\kappa\rho
Q}{\pi^{2}R^{4}}+\frac{\rho\alpha Q^{2}}{\pi^{3}R^{6}}\left|\frac{\partial A}{\partial
x}\right|\right)
\end{equation}

\begin{equation}
|\tau_{w}|=\left(\frac{\kappa\rho Q}{2\pi^{2}R^{3}}+\frac{\rho\alpha
Q^{2}}{2\pi^{3}R^{5}}\frac{\partial A}{\partial R}\left|\frac{\partial R}{\partial x}\right|\right)
\end{equation}

\begin{equation}
|\tau_{w}|=\left(\frac{\kappa\rho Q}{2\pi^{2}R^{3}}+\frac{\rho\alpha Q^{2}}{2\pi^{3}R^{5}}2\pi
R\left|\frac{\partial R}{\partial x}\right|\right)
\end{equation}

\begin{equation}
|\tau_{w}|=\left(\frac{\kappa\rho Q}{2\pi^{2}R^{3}}+\frac{\rho\alpha
Q^{2}}{\pi^{2}R^{4}}\left|\frac{\partial R}{\partial x}\right|\right)
\end{equation}
that is

\begin{equation}
|\tau_{w}|=\frac{B}{R^{3}}+\frac{C}{R^{4}}\left|\frac{\partial R}{\partial x}\right|
\end{equation}
where

\begin{equation}
B=\frac{\kappa\rho Q}{2\pi^{2}}>0\,\,\,\,\,\,\,\,\,\,\ \textrm{and} \,\,\,\,\,\,\,\,\,\,\,
C=\frac{\rho\alpha Q^{2}}{\pi^{2}}>0
\end{equation}

In the following we will show that $f\equiv|\tau_{w}|$ is a monotonically increasing function of
$x$ by demonstrating that the first derivative of $f$ is positive, i.e. $\frac{\partial f}{\partial
x}>0$.

\begin{equation}
\frac{\partial f}{\partial x}=\frac{\partial}{\partial
x}\left(\frac{B}{R^{3}}+\frac{C}{R^{4}}\left|\frac{\partial R}{\partial x}\right|\right)
\end{equation}

Now, since $R$ is a monotonically decreasing function of $x$ we have

\begin{equation}
\left|\frac{\partial R}{\partial x}\right|=-\frac{\partial R}{\partial x}
\end{equation}
and

\begin{equation}
\frac{\partial\left|\frac{\partial R}{\partial x}\right|}{\partial
x}=-\frac{\partial^{2}R}{\partial x^{2}}
\end{equation}

Therefore

\begin{equation}
\frac{\partial f}{\partial x}=-\frac{3B}{R^{4}}\frac{\partial R}{\partial
x}-\frac{4C}{R^{5}}\frac{\partial R}{\partial x}\left|\frac{\partial R}{\partial
x}\right|-\frac{C}{R^{4}}\frac{\partial^{2}R}{\partial x^{2}}
\end{equation}

\begin{equation}
\frac{\partial f}{\partial x}=\frac{3B}{R^{4}}\left|\frac{\partial R}{\partial
x}\right|+\frac{4C}{R^{5}}\left|\frac{\partial R}{\partial
x}\right|^{2}-\frac{C}{R^{4}}\frac{\partial^{2}R}{\partial x^{2}}
\end{equation}

\begin{equation}
\frac{\partial f}{\partial
x}>0\,\,\,\,\Rightarrow\,\,\,\,\frac{C}{R^{4}}\frac{\partial^{2}R}{\partial
x^{2}}<\frac{3B}{R^{4}}\left|\frac{\partial R}{\partial
x}\right|+\frac{4C}{R^{5}}\left|\frac{\partial R}{\partial x}\right|^{2}
\end{equation}
i.e.

\begin{equation}
\frac{\partial^{2}R}{\partial x^{2}}<\frac{3B}{C}\left|\frac{\partial R}{\partial
x}\right|+\frac{4}{R}\left|\frac{\partial R}{\partial
x}\right|^{2}\,\,\,\,\,\,\,\,\left(R>0\right)\label{curvCond}
\end{equation}

Now, $\frac{\partial^{2}R}{\partial x^{2}}$ is either negative or positive. If it is negative then
the condition given by Equation \ref{curvCond} is always satisfied because the right hand side is
strictly positive. However, if it is positive then since

\begin{equation}
\frac{B}{C}=\frac{\kappa}{2\alpha Q}
\end{equation}
we can show that in all practical situations the first term on the right hand side is greater than
the left hand side regardless of the magnitude of the second term on the right hand side. The
reason is that prior to yield $Q$ is infinitesimally small, and therefore $\frac{B}{C}$ is very
large in magnitude and positive in sign. Hence for any tangible pressure drop that makes
$\left|\frac{\partial R}{\partial x}\right|$ finite, which is automatically satisfied for any
tangible yield-stress, the first term on the right hand side is very large. Therefore, for all
practical purposes where the curvature of the tube, as quantified by the second derivative of $R$
with respect to $x$, is constrained within the physical limits the condition given by Equation
\ref{curvCond} will be satisfied. This means that in both cases the first derivative of $f$ is
greater than zero and hence the absolute value of the wall shear stress, $|\tau_{w}|$, is a
monotonically increasing function of the tube axial coordinate $x$.

As a result, the bottleneck according to the highly-viscous fluid approach will always be at the
inlet boundary. This finding is confirmed by numerical experiments where all the results indicate
that the bottleneck is at the inlet boundary. It should be remarked that it can be demonstrated
that for the one-dimensional Navier-Stokes flow in an elastic tube whose mechanical response is
described by Equation \ref{pAEq2} it is always the case that $\frac{\partial^{2}R}{\partial
x^{2}}<0$, i.e. the axial dependency of the radius, area and pressure concave downward (refer for
instance to Figure \ref{pAnalFig}). However, we will not prove this here to avoid unnecessary
details because, as demonstrated already, this is not needed for establishing our case.

To sum up, the identification of the bottleneck through the inspection of the axial pressure field
by using the numerical or the analytical solution is the generally valid method and hence it should
provide a definite and reliable answer for determining the yield condition. However, as
demonstrated in the previous paragraphs, at least for all the practical purposes in which the
distensible tube is not subjected to extremely large pressure boundary conditions which invalidate
the pressure-area relation, the bottleneck and hence the yield condition can be identified from
just inspecting the magnitude of the wall shear stress at the inlet boundary, which as explained
earlier only requires computing the one-sided pressure gradient numerically or analytically. In all
cases, tests can be carried out to verify that the bottleneck is actually at the inlet boundary.

%XXXXXXXXXXXXXXXXXXXXXXXXXXXXXXXXXXXXXXXXXXXXXXXXXXXXXXXXXXXXXXXXX
\section{General Assessment}

The method proposed in this paper for the identification of the yield condition in the mobilization
of \ys\ materials in distensible tubes is based on a number of simplifying assumptions regarding
the fluid, flow and tube. The purpose of this investigation is to derive the yield condition from
logical reasoning through the employment of the principles of mechanics, rheology and fluid
dynamics. Many practical factors that usually arise in real life can change this condition and
hence require a deeper inspection to identify the yield condition more rigorously. The real \ys\
flow systems are more physically complex than the description provided by our model or any other
model since all these models are based on mathematical idealizations and physical simplifications.
However, we believe this investigation can provide the ground for further investigations in the
future through which more complex factors can be incorporated in the modeled \ys\ flow systems to
provide better predictions for the yield condition.

%XXXXXXXXXXXXXXXXXXXXXXXXXXXXXXXXXXXXXXXXXXXXXXXXXXXXXXXXXXXXXXXXX
\section{Conclusions}

The investigation in the present paper lead to the proposal of a method for identifying the yield
condition in the mobilization of \ys\ materials through deformable cylindrical tubes with a
constant unstressed cross sectional area along their axial length. Two possible scenarios for
modeling the \ys\ materials prior to yield have been considered: a solid-like approach and a
highly-viscous fluid approach. The logical consequences of these two approaches were derived using
the principles of rheology and fluid dynamics as summarized in the rheological characteristics of
the deformed material, Navier-Stokes system and the distensibility characteristics of the tube
wall. General mechanical principles were also employed in this investigation with regard to the
solid-like approach.

Although the derived yield condition is based on several simplifying assumptions with regard to the
type of flow, fluid and tube characteristics and hence it applies to a rather idealized \ys\ flow
system, the proposed model can serve as a first step for more elaborate \ys\ models that
incorporate more physical factors that influence the yield condition.

In the course of this investigation, an analytical expression linking the axial pressure field
inside a compliant tube having elastic properties to its axial coordinate along its length has also
been derived and validated by a residual-based numerical method from the lubrication theory.

\clearpage
%XXXXXXXXXXXXXXXXXXXXXXXXXXXXXXXXXXXXXXXXXXXXXXXXXXXXXXXXXXXXXXXXXXX
\phantomsection \addcontentsline{toc}{section}{Nomenclature} %
{\noindent \LARGE \bf Nomenclature} \vspace{0.5cm}

\begin{supertabular}{ll}
$\alpha$                &   correction factor for axial momentum flux \\
$\beta$                 &   stiffness coefficient in pressure-area relation \\
%$\gamma$                &   stiffness parameter in the first pressure-area relation \\
%$\delta$                &   Murray's law index \\
$\kappa$                &   viscosity friction coefficient \\
$\mu$                   &   fluid dynamic viscosity \\
%$\nu$                   &   fluid kinematic viscosity \\
$\rho$                  &   fluid mass density \\
$\varsigma$             &   Poisson's ratio of tube wall \\
$\ysS$                  & yield-stress \\
$\wsS$                  & shear stress at tube wall \\
%$\boldsymbol{\omega}$   &   vector of test functions in the weak form of finite element \\
%$\Omega$                &   solution domain \\
%$\Omega_e$              &   solution domain of an element \\
%$\partial \Omega$       &   boundary of the solution domain \\
\\
$A$                     &   tube cross sectional area \\
%$A_{BC}$                &   boundary condition for vessel cross sectional area \\
$A_{in}$                &   tube cross sectional area at inlet \\
$A_o$                   &   tube cross sectional area at reference pressure \\
$A_{ou}$                &   tube cross sectional area at outlet \\
$A_{\parallel}$         & luminal area parallel to tube axis \\
%$\mathbf{B}$            &   matrix of force terms in the 1D Navier-Stokes equations \\
%$\mathbf{C}$            &  conductance matrix \\
$E$                     &   Young's elastic modulus of tube wall \\
$F_{\perp}$             & normal force \\
%$f$                     &   flow continuity residual function \\
%$\mathbf{F}$            &   flux matrix in the 1D Navier-Stokes equations \\
%$h$                     &   length of element \\
$h_o$                   &   tube wall thickness at reference pressure \\
%$\mathbf{H}$            &   matrix of partial derivative of $\mathbf{F}$ with respect to $\mathbf{U}$ \\
%$\mathbf{J}$            &   Jacobian matrix \\
%$l^T_{1,2}$             &   left eigenvalues of $\mathbf{H}$ matrix (?) \\
$L$                     &   tube length \\
%$n$                     &   number of network nodes \\
%$\mathfrak{N}$          &   norm of residual vector \\
$p$                     &   pressure \\
%$\mathbf{p}$            &  pressure vector \\
$p_{in}$                &  inlet boundary pressure \\
$p_{ou}$                &  outlet boundary pressure \\
$\delta p$              &   infinitesimal pressure drop \\
$\Delta P$              &   pressure drop across the tube \\
%$\Delta\mathbf{p}$      &   pressure perturbation vector \\
%$p$                     &   order of interpolating polynomial \\
%$p_A$                   &   amplitude of sinusoidal input pressure signal \\
%$p_{o}$                   &   reference pressure (?) \\
%$\Delta p$              &   pressure step \\
%$q$                     &   dummy index for quadrature point \\
$Q$                     &   volumetric flow rate \\
%$\mathbf{Q}$            &  total flow column vector \\
%$Q_N$                    &   volumetric flow rate of 1D Navier-Stokes model \\
%$Q_P$                    &   volumetric flow rate of \pois\ model \\
%$Q_{BC}$                &   boundary condition for volumetric flow rate \\
%$r$                     &   flow continuity residual \\
%$\mathbf{r}$             &  residual vector \\
$R$                     &   tube radius \\
%$r$                     &   tube radius \\
%$r_{d}$                 &   radius of daughter tube \\
%$r_{m}$                 &   radius of mother tube \\
%$\mathbf{R}$            &   weak form of residual vector \\
%$S_{a}$                 &   analytic solution \\
%$S_{n}$                 &   numeric solution \\
$t$                     &   time \\
%$\Delta t$              &   time step \\
%$u$                     &   local axial speed of fluid \\
%$\overline{u}$          &   mean axial speed of fluid \\
%$\mathbf{U}$            &   vector of finite element variables \\
%$\mathbf{\Delta U}$     &   vector of change in $\mathbf{U}$ \\
$x$                     &   tube axial coordinate \\
$\delta x$              &   infinitesimal length along tube axis \\
%\\
%1D                      &   one-dimensional \\
%FE                      &   finite element \\
%NS                      &   Navier-Stokes \\
%TD                      &   time-dependent \\
%TI                      &   time-independent \\

\end{supertabular}

\clearpage
\phantomsection \addcontentsline{toc}{section}{References} %
\bibliographystyle{unsrt}
%\bibliography{Bibl}

\end{document}